\documentclass[12pt]{article}

\usepackage{amsmath}
\usepackage{amsthm}
\usepackage{amsfonts}
\usepackage{amssymb}
\usepackage{latexsym} 
\usepackage{epsfig}
\usepackage{graphicx}
\usepackage{calc}  %%required to make lists work correctly
\usepackage[matrix,tips,graph,curve,web]{xy}

\makeatletter

\def\@maketitle{%
  \newpage
  \null
  \let \footnote \thanks
    {\normalfont\sffamily\bfseries\Large\noindent\@title \par}%
    \vskip 1em%
    {\normalfont\sffamily %\large
        \noindent
        \@author
        \par}
  \par
  \vskip 4em}
\def\@seccntformat#1{\csname the#1\endcsname{.\ }}
\renewcommand\section{\@startsection {section}{1}{\z@}%
                                   {-3.0ex \@plus -1ex \@minus -.2ex}%
                                   {1.5ex \@plus.2ex}%
                                   {\normalfont\large\bfseries}}
\renewcommand\subsection{\@startsection{subsection}{2}{\z@}%
                                     {-2.75ex\@plus -1ex \@minus -.2ex}%
                                     {1.5ex \@plus .2ex}%
                                   {\normalfont\large}}
\def\fnum@figure{\normalfont\footnotesize\figurename~\thefigure}
\renewcommand\tableofcontents{%
    \section*{\contentsname
        \@mkboth{%
           \MakeUppercase\contentsname}{\MakeUppercase\contentsname}}%
    \@starttoc{toc}%
    }
\renewcommand*\l@part[2]{%
  \ifnum \c@tocdepth >-2\relax
    \addpenalty\@secpenalty
    \addvspace{2.25em \@plus\p@}%
    \begingroup
      \setlength\@tempdima{3em}%
      \parindent \z@ \rightskip \@pnumwidth
      \parfillskip -\@pnumwidth
      {\leavevmode
       \large \bfseries #1\hfil \hb@xt@\@pnumwidth{\hss #2}}\par
       \nobreak
       \if@compatibility
         \global\@nobreaktrue
         \everypar{\global\@nobreakfalse\everypar{}}%
      \fi
    \endgroup
  \fi}
\renewcommand*\l@section[2]{%
  \ifnum \c@tocdepth >\z@
    \addpenalty\@secpenalty
    \addvspace{1.0em \@plus\p@}%
    \setlength\@tempdima{1.5em}%
    \begingroup
      \parindent \z@ \rightskip \@pnumwidth
      \parfillskip -\@pnumwidth
      \leavevmode \sffamily\bfseries
      \advance\leftskip\@tempdima
      \hskip -\leftskip
      #1\nobreak\hfil \nobreak\hb@xt@\@pnumwidth{\hss #2}\par
    \endgroup
  \fi}
\renewcommand*\l@subsection{\sffamily\@dottedtocline{2}{1.5em}{2.3em}}
\renewcommand*\l@subsubsection{\@dottedtocline{3}{3.8em}{3.2em}}
\renewcommand*\l@paragraph{\@dottedtocline{4}{7.0em}{4.1em}}
\renewcommand*\l@subparagraph{\@dottedtocline{5}{10em}{5em}}
\makeatother

\makeatletter 
\@addtoreset{equation}{section}
\makeatother

\renewcommand{\theequation}{\thesection.\arabic{equation}}

\theoremstyle{plain}
\newtheorem{theorem}[equation]{Theorem}
\newtheorem{corollary}[equation]{Corollary}
\newtheorem{lemma}[equation]{Lemma}

\theoremstyle{definition}
\newtheorem{definition}[equation]{Definition}

  {\begin{list}{}%
    {%
    \settowidth{\labelwidth}{#1}%
    \setlength{\itemindent}{0pt}%
    \setlength{\labelsep}{1em}%
    \setlength{\leftmargin}{\labelwidth+\parindent+\labelsep}%
    \setlength{\itemsep}{0pt}%
    \setlength{\parsep}{.6ex}}}%
  {\end{list}}

  {\begin{list}{}%
    {%
    \settowidth{\labelwidth}{#1}%
    \setlength{\itemindent}{0pt}%
    \setlength{\labelsep}{1em}%
    \setlength{\leftmargin}{\labelwidth+\labelsep}%
    \setlength{\itemsep}{0pt}%
    \setlength{\parsep}{.6ex}}}%
  {\end{list}}

\makeatletter
\newenvironment{subequations*}{% same thing but without incrementing
  \begingroup % conservative approach
  \let\protect\@nx
  \edef\@tempa{\def\@nx\theparentequation{\theequation}}%
  \@xp\endgroup\@tempa
  \setcounter{parentequation}{\value{equation}}%
  \setcounter{equation}{0}%
  \def\theequation{\theparentequation\alph{equation}}%
  \ignorespaces
}{%
  \setcounter{equation}{\value{parentequation}}%
  \global\@ignoretrue
}

\renewcommand\det{{\rm det\,}}

\def\d/{/\mspace{-6.0mu}/}

\usepackage{simplemargins}
\usepackage{amssymb,amsfonts,amsmath,amsthm,url}

\theoremstyle{definition}
\def\od{\stackrel{\mathrm{def}}{=}}

\def\RR{\mathbb{R}}
\def\supp{\operatorname{supp}}

\def\det{\operatorname{det}}

\def\FP{\operatorname{FP}}
\def\tilG{\widetilde{G}}
\def\Wtil{\widetilde{W}}

\usepackage{multicol}
\usepackage{color}
\definecolor{gold}{rgb}{0.85,.66,0}
\definecolor{cherry}{rgb}{0.9,.1,.2}
\definecolor{burgundy}{rgb}{0.8,.2,.2}
\definecolor{orangered}{rgb}{0.85,.3,0}
\definecolor{orange}{rgb}{0.85,.4,0}
\definecolor{olive}{rgb}{.45,.4,0}
\definecolor{lime}{rgb}{.6,.9,0}
\definecolor{green}{rgb}{.2,.7,0}
\definecolor{darkgreen}{rgb}{.1,.5,0}
\definecolor{grey}{rgb}{.4,.4,.2}
\definecolor{brown}{rgb}{.4,.2,.1}
\definecolor{blue}{rgb}{0,.0, .81}
\definecolor{bluepurple}{rgb}{.3, .0, .7}

\begin{document}
%\begin{center}
\noindent {\large \textbf{On graphical domination for threshold-linear networks with recurrent excitation and global inhibition}}\\
%\end{center}
\noindent Carina Curto\\
{\it October 6, 2025}

\section*{Abstract}
Graphical domination was first introduced in \cite{fp-paper} in the context of combinatorial threshold-linear networks (CTLNs). There it was shown that when a domination relationship exists between a pair of vertices in a graph, certain fixed points in the corresponding CTLN can be ruled out. Here we prove two new theorems about graphical domination, and show that they apply to a significantly more general class of recurrent networks called generalized CTLNs (gCTLNs). Theorem 1 establishes that if a dominated node is removed from a network, the reduced network has exactly the same fixed points. Theorem 2 tells us that by iteratively removing dominated nodes from an initial graph $G$, the final (irreducible) graph $\tilG$ is unique. We also introduce another new family of TLNs, called E-I TLNs, consisting of $n$ excitatory nodes and a single inhibitory node providing global inhibition. We provide a concrete mapping between the parameters of gCTLNs and E-I TLNs built from the same graph such that corresponding networks have the same fixed points. We also show that Theorems 1 and 2 apply equally well to E-I TLNs, and that the dynamics of gCTLNs and E-I TLNs with the same underlying graph $G$ exhibit similar behavior that is well predicted by the fixed points of the reduced graph $\tilG$.

%\cite{Notices, extended-notices, CTLN-diversity, book-chapter, fp-paper, rule-of-thumb, stable-fp-paper, n5-github, lienkaemper2025}

\tableofcontents

% Gaudi matrix
%W = \left(\begin{array}{cccccc} 
%0   & 0   & 0   & a_4 & a_5 & -1 \\
%a_1 & 0   & 0   & 0   & a_5 & -1 \\
%a_1 & a_2 & 0   & 0   & 0   & -1 \\
%0   & a_2 & a_3 & 0   & 0   & -1 \\
%0   & 0   & a_3 & a_4 & 0   & -1 \\
%c_1 & c_2 & c_3 & c_4 & c_5 & 0 
%\end{array}\right)

% baby chaos matrix
%W = \left(\begin{array}{cccccc} 
%0   & 0   & 0   & 0   & a_5 & -1 \\
%a_1 & 0   & a_3 & 0   & 0   & -1 \\
%0   & 0   & 0   & 0   & a_5 & -1 \\
%a_1 & 0   & a_3 & 0   & 0   & -1 \\
%0   & a_2 & 0   & a_4 & 0   & -1 \\
%c_1 & c_2 & c_3 & c_4 & c_5 & 0 
%\end{array}\right)

\section{Introduction: basic definitions and summary of results}
Graphical domination is a relationship that a pair of nodes can have in a simple directed graph $G$.\footnote{{\it Simple} means that there are no multi-edges and no self loops.} It was first introduced in \cite{fp-paper} in the context of combinatorial threshold-linear networks (CTLNs), for modeling recurrent networks in neuroscience, but the definition itself is entirely about graphs.

\begin{definition}\label{def:dom}
Let $j,k \in [n]$ be vertices of $G$. We say that $k$ {\em graphically dominates} $j$ in $G$, and write $k > j$, if the following two conditions hold:
\begin{itemize}
\item[(i)] For each vertex $i \in [n]\setminus\{j,k\}$, if $i \to j$ then $i \to k$.
\item[(ii)] $j \to k$ and $k \not\to j$.
\end{itemize}
If there exists a $k$ that graphically dominates $j$, we say that $j$ is a {\it dominated node} (or {\it dominated vertex}) of $G$. If $G$ has no dominated nodes, we say that it is {\it domination free}.
\end{definition}

Note that graphical domination is defined purely in terms of the graph, without reference to any associated network or dynamical system. However, the reason we were originally interested in domination is that it gave us constraints on the sets of fixed points of combinatorial threshold-linear networks (CTLNs) \cite{fp-paper, Notices}. Our original definition of domination involved variants with respect to different subsets of the nodes of the graph, $\sigma \subseteq [n]$, and the different cases (inside-in, inside-out, outside-in, outside-out) allowed us to rule and rule out various $\sigma$ from being fixed point supports of the network \cite{fp-paper, Notices, extended-notices}. These more technical results were needed because, at the time, we did not know that a dominated node could be removed from the network without altering the fixed points. 

In this work, we prove two new theorems about domination that are significantly more powerful than our previous results. In particular, we no longer need the various variants of domination with respect to subsets; we only need the simplest variant with respect to the full graph $G$, given in Definition~\ref{def:dom} above. Our new theorems also apply to a much wider class of recurrent networks called {\it generalized} CTLNs (gCTLNs), as well as to a corresponding family of {\it excitatory-inhibitory} TLNs (E-I TLNs). We define both of these new families here, beginning with gCTLNs.

\paragraph{gCTLNs.} 
A threshold-linear network (TLN) has dynamics that are given by the standard TLN equations:
\begin{eqnarray}
\tau_i\dfrac{dx_i}{dt} = -x_i + \left[ \sum_{j=1}^n W_{ij} x_j + b_i \right]_+,
\end{eqnarray}
where $W$ is an $n\times n$ matrix with real-valued entries, $b \in \RR^n$, and $\tau_i > 0$ is the timescale for each neuron, and $[z]_+ = \max\{z,0\}$ (the standard ReLU nonlinearity). Such a network is fully specified by the parameters $(W,b,\tau_i)$. When we set $\tau_i = 1$ for all neurons, so that time is measured in units of a single timescale, we simplify the notation to $(W,b)$.

We call such a network a gCTLN if it is constructed from a directed graph $G$, according to the following rule:
\begin{eqnarray}\label{eq:Wmtx}
W_{ij} = \left\{\begin{array}{cc} -1+\varepsilon_j & \text{ if } j \to i,\\
-1-\delta_j &  \text{ if } j \not\to i, \\
0 & \text{ if } i = j.\end{array}\right.
\end{eqnarray}
Here we assume $G$ has $n$ vertices and $W$ is an $n \times n$ real-valued matrix.
The parameters $\varepsilon_j, \delta_j > 0$ can be different for each node, and $\varepsilon_j < 1$ so that we guarantee that $W_{ij} \leq 0$.
Moreover, the vector $b \in \RR^n$ is defined as $b_i = \theta > 0$ for all $i \in [n]$, and the timescales are all taken to be equal and set to $\tau_i = 1$. The data for a gCTLN is thus completely specified by a directed graph $G$ and the $2n+1$ parameters: $\{\varepsilon_j, \delta_j, \theta\}_{j \in [n]}$.

For example, if $G$ is the graph with $n = 3$ nodes and four edges, $1 \leftrightarrow 2 \to 3 \to 1$, the corresponding gCTLN with parameters $\{\varepsilon_j, \delta_j, \theta\}$ has weight matrix $W$ given by:
$$W = \left(\begin{array}{ccc} 
0 & -1+\varepsilon_2 & -1+\varepsilon_3 \\
-1+\varepsilon_1 & 0 & -1-\delta_3 \\
-1-\delta_1 & -1+\varepsilon_2 & 0 \\
 \end{array}\right)$$

The definition of gCTLNs is very similar in spirit to our definition of CTLNs from prior work \cite{fp-paper, Notices,CTLN-diversity}. However, CTLNs only have three parameters: $\varepsilon, \delta$ and $\theta$. They are the special case of gCTLNs where $\varepsilon_j = \varepsilon$ and $\delta_j = \delta$ for all $j = 1,\ldots,n$. 

For a given choice of $\{\varepsilon_j, \delta_j, \theta\}$, the dynamics of the associated gCTLN are fully determined by the graph $G$. Like all TLNs, these networks have a collection of fixed points in $\RR^n$ at which the vector field $(dx_1/dt, \ldots, dx_n/dt)$ vanishes. These fixed points necessarily lie in the positive orthant, $\RR_{\geq 0}^n$, and can be labeled by their supports. We use the notation $\FP(G) = \FP(G,\varepsilon_j,\delta_j)$ to denote the fixed point supports of a gCTLN with a given set of parameters:
$$\FP(G) \od \{\sigma \subseteq [n] \mid \text{ the gCTLN has a fixed point } x^* \text{ with } \supp(x^*) = \sigma \}.$$
 Note that $\theta$ merely rescales the fixed points, and does not alter their supports, so we do not include it in the notation. 
 
\begin{figure}[!h]
\begin{center}
\includegraphics[width=5.5in]{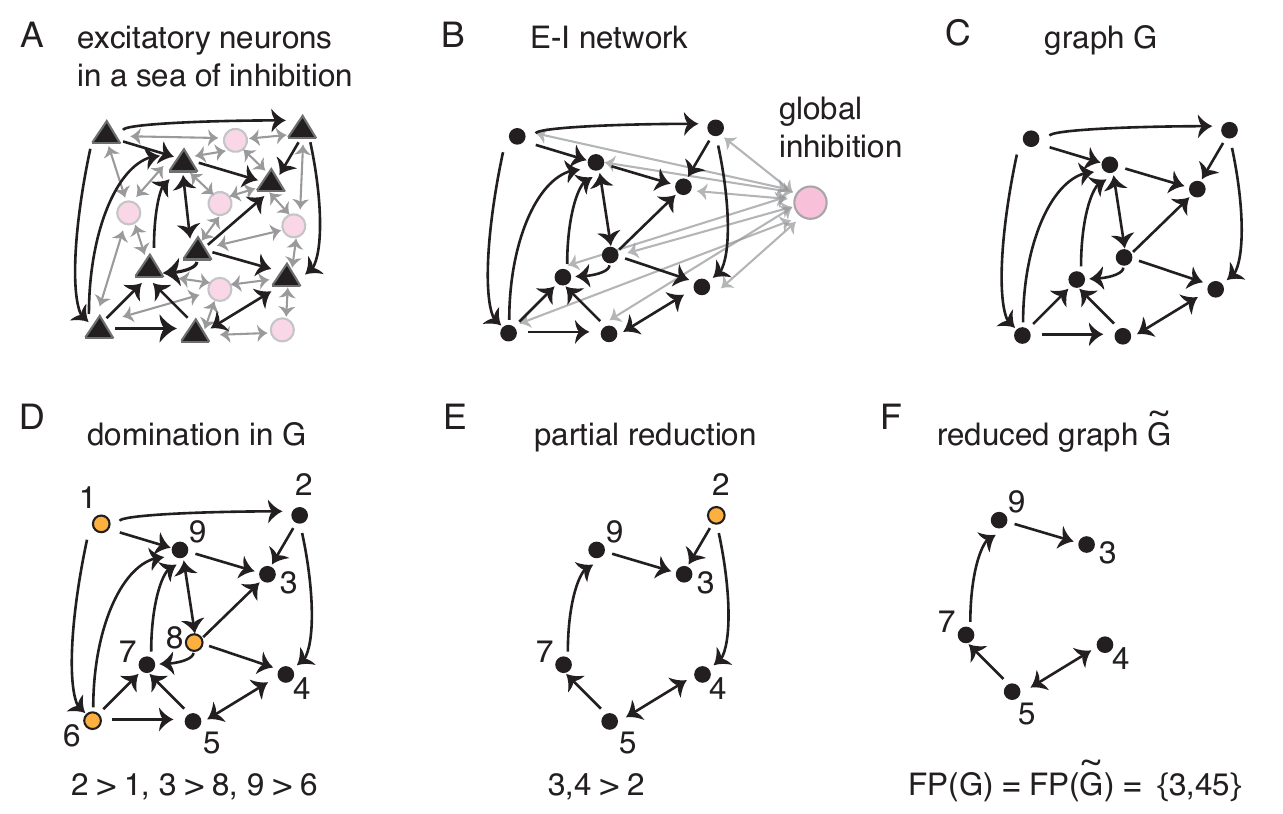}
\end{center}
\caption{Graph-based networks and graphical domination.}
\label{fig:net+domination}
\end{figure}

 Although the interactions $W_{ij}$ are all inhibitory, we think of this network as modeling excitatory neurons in a sea of global inhibition. In other words, an edge $j \to i$ in the graph corresponds to a sum of excitation and global inhibition, leading to an effective weak inhibition of weight $W_{ij} = -1+\varepsilon_j$. When there is no edge, the inhibition remains strong. The graph $G$ thus captures the pattern of weak and strong inhibition in an effectively inhibitory network (Figure~\ref{fig:net+domination}A-C).

\paragraph{Domination theorems.}
Our first new domination theorem states that a dominated node $j$ can be removed from a gCTLN (and thus a CTLN) without altering the fixed points of the network dynamics.

\begin{theorem}[Theorem 1] \label{thm:domination}
Suppose $j$ is a dominated node in a directed graph $G$.
Then the fixed points of a gCTLN constructed from $G$ satisfy 
$$\FP(G) = \FP(G|_{[n]\setminus j}),$$
for any choice of gCTLN parameters $\{\varepsilon_i, \delta_i, \theta\}$.
\end{theorem}

This theorem also holds for E-I TLNs, which will be defined in the next section. The proof, given in Section~\ref{sec:thm1-proof}, treats both cases in parallel.

Theorem 1 can be applied iteratively, as new nodes can become dominated in the subgraph
$G|_{[n]\setminus j}$ that remains after the removal of the initial dominated node. Continuing in this manner, we can always reduce our graph $G$ down to a subgraph $\tilG$ that is domination free. 

\begin{definition}\label{def:reduced}
Let $G$ be a directed graph on $n$ nodes. We say that $\tilG$ is the {\it reduced graph of} $G$ if the following conditions hold:
\begin{enumerate}
\item $\tilG$ is an induced subgraph of $G$, so that $\tilG = G|_\tau$, for some $\tau \subseteq [n]$,
\item $\tilG$ is domination free, and
\item $\tilG$ can be obtained from $G$ by iteratively removing dominated nodes.
\end{enumerate}
\end{definition}

For example, in the graph of Figure~\ref{fig:net+domination}D, we see that node 2 dominates 1, 3 dominates 8, and 9 dominates 6. We can thus remove (in any order) all three of the dominated nodes 1, 6, and 8. In the reduced graph, 
Figure~\ref{fig:net+domination}E, we now see that 2 is a dominated node, even though it was not dominated in the original graph. The final reduced graph is given in Figure~\ref{fig:net+domination}F. 
% Fix FP(G) in Figure 1F! It's wrong.

Our second domination theorem states that the reduced graph $\tilG$ is unique. In other words, it does not matter in what order dominated nodes are removed from $G$ -- the removal process will always terminate in the same graph.

\begin{theorem}[Theorem 2] \label{thm:dom-uniqueness}
Let $G$ be a directed graph on $n$ nodes. Then the reduced graph $\tilG$ is unique. 
\end{theorem}

\noindent The proof is given in Section~\ref{sec:uniqueness}.

As a corollary of Theorems 1 and 2, we have the following very useful fact.

\begin{corollary} Let $G$ be a directed graph, and let $\tilG$ be its unique domination-free reduction. Then for any gCTLN parameters $\{\varepsilon_i, \delta_i, \theta\}$, appropriately restricted to $\tilG$, we have:
$$\FP(G) = \FP(\tilG).$$
\end{corollary}

As we will later see, this corollary will also hold for E-I TLNs since Theorems 1 and 2 apply to these networks as well.

The reason these results are especially powerful is that in competitive TLNs, the activity tends to flow in the direction of the fixed points, even if they are all unstable. We do not have a precise formulation or proof of this observation, but the intuition coming from computational experiments and theoretical considerations is strong. In particular, the Perron-Frobenius theorem guarantees that every unstable fixed point is a saddle, with at least one attracting direction (and this direction corresponds to the largest, and most negative, eigenvalue) \cite{Notices}. The Parity theorem \cite{CTLN-diversity} also suggests that attractors must live ``near'' the fixed points. In particular, any convex attracting set must contain at least one fixed point. Overall, we expect the dynamics of a gCTLN constructed from a graph $G$ to flow towards attractors that are concentrated on the nodes of $\tilG$.

Although this intuition for the network dynamics stems from observations about fixed points and attractors in CTLNs \cite{rule-of-thumb, Notices, CTLN-diversity}, which naturally extend to gCTLNs, we will see in the next section that it also applies to a new family of {\it excitatory-inhibitory TLNs} (E-I TLNs) which are not competitive and where the Perron-Frobenius theorem does not hold. Nevertheless, we find that there is a mapping between E-I TLNs and gCTLNs such that the fixed points match (see Theorem 3), and the asymptotic behavior appears to be nearly identical when $\tau_I \ll 1$ (setting the excitatory timescale $\tau_E = 1$). In particular, we show that the domination theorems apply equally well to E-I TLNs.

\paragraph{Roadmap.} The rest of this paper is organized as follows. In Section~\ref{sec:E-I}, we define E-I TLNs and give an explicit mapping between the parameters of an E-I TLN and the corresponding gCTLN with matching fixed points. We also provide a number of examples to show that the dynamics of E-I TLNs and gCTLNs closely match well beyond the fixed points. In Section~\ref{sec:dom-proofs}, we prove the two new domination theorems, Theorem 1 and Theorem 2. 

\section{E-I TLNs}\label{sec:E-I}
In this section, we define a family of excitatory-inhibitory TLNs from a directed graph according to a prescription that is similar in spirit to the definition of gCTLNs. We will then show that there is a mapping between such a network and a corresponding gCTLN with matching fixed point structure. This construction is very similar to the E-I networks defined in \cite{lienkaemper2025}, corresponding to CTLNs. In that work, the E-I networks served as a stepping stone connecting CTLNs to larger stochastic spiking networks with probabilistic connections between populations. Here we simply introduce them as a companion family to the gCTLNs with similar dynamics, despite having fundamentally different $W$ matrices. In particular, we will show that our new domination theorems apply equally well to excitatory-inhibitory networks, without needing the special competitive conditions of gCTLNs (see Theorem~\ref{thm:domination-both} in the next section).

Given any directed graph $G$ on $n$ vertices, we construct an excitatory-inhibitory threshold-linear network (E-I TLN) with dynamics given by:
\begin{eqnarray}\label{eq:E-I-network}
\dfrac{dx_i}{dt} &=& -x_i + \left[ \sum_{j=1}^n W_{ij} x_j + W_{iI}x_I + b_i \right]_+, \; i=1,\ldots,n, \\
\tau_I\dfrac{dx_I}{dt} &=& -x_I + \left[ \sum_{j=1}^n W_{Ij} x_j + b_I \right]_+
\end{eqnarray}
The threshold-linear function $[z]_+ = \max\{z,0\}$ is the standard ReLU nonlinearity.
The connectivity matrix $W$ is defined from $G$ as follows. For $i,j = 1,\ldots,n$ (the ``E'' nodes), we have excitatory weights $a_j > 0$ which depend only on the pre-synaptic node:
\begin{eqnarray}\label{eq:Wmtx}
W_{ij} = \left\{\begin{array}{cc} a_j & \text{ if } j \to i \text{ in } G,\\
0 &  \text{ if } j \not\to i  \text{ in } G. \end{array}\right.
\end{eqnarray}
The vertices of $G$ correspond to excitatory neurons, and the weights $W_{ij}$ between them are all nonnegative.

Additionally, the weights to and from the inhibitory node $I = n+1$ are given as follows:
$$W_{Ij} = c_j, \; W_{iI} = -1, \text{ and } W_{II} = 0,   \;\; \text{for} \; i,j=1,\ldots,n.$$
Note that the inhibitory node does not correspond to any of the vertices in $G$. Rather, it has all-to-all connections with each excitatory node, so there is no graphical information to encode.

The network also includes self-excitation terms,
$$W_{ii} = c_i,$$
which are meant to precisely cancel the self-inhibition that stems from $x_i$'s contribution to the steady-state value of the inhibitory node $x_I$, which is $W_{Ii}x_i$. Alternatively, we could rewrite the excitatory equations as:
$$\dfrac{dx_i}{dt} = -x_i + \left[ \sum_{j=1}^n W_{ij} x_j + W_{iI}(x_I - W_{Ii} x_i) + b_i \right]_+, \; i=1,\ldots,n,$$
so that only the inhibition coming from the other nodes, $x_I - W_{Ii} x_i$, feeds back into the $x_i$ equation. This turns out to be equivalent to defining $W_{ii} = -W_{iI}W_{Ii} = c_i$. In order to keep the TLN equations in the same form as before, it is more convenient to keep the simpler inhibitory interaction terms, $W_{iI} x_I$, and add the self-excitation.

Unless otherwise specified, we will set the vector $b \in \RR^{n+1}$ to be $b_i = \theta > 0$ for all $i = 1,\ldots, n$, as in gCTLNs, and $b_I = b_{n+1} = 0$ for the inhibitory node.
Finally, we will require the parameters satisfy:
$$a_j > 0 \;\; \text{and} \;\; 1 < c_j < 1+a_j.$$
This is equivalent to the requirement that $\varepsilon_j,\delta_j > 0$ for the corresponding gCTLN.

An E-I TLN is fully specified by a directed graph $G$ and the parameters $\{a_j,c_j,\theta,\tau_I\}$. The inhibitory timescale, $\tau_I$, is presumed to be smaller than the excitatory timescale, $\tau_E$, which has been implicitly set to equal $1$ (i.e., we measure time in units of $\tau_E$). 

For example, the graph with $n = 3$ nodes and four edges, $1 \leftrightarrow 2 \to 3 \to 1$, yields the following $4 \times 4$ weight matrix, where the index $n+1 = 4$ corresponds to the inhibitory node.
$$W = \left(\begin{array}{cccc} 
c_1 & a_2 & a_3 & -1 \\
a_1 & c_2 & 0 & -1 \\
0 & a_2 & c_3 & -1 \\
c_1 & c_2 & c_3 & 0 \end{array}\right)$$
For larger graphs, the sparsity of the $W$ matrix in an E-I TLN will reflect the sparsity of the graph, because $W_{ij} = 0$ when $j \not\to i$ in $G$. This is different from the case of CTLNs and gCTLNs, where the matrices are always dense since missing edges in the graph correspond to strongly inhibitory (nonzero) weights.

\subsection{Domination in E-I TLNs}
\begin{figure}
\begin{center}
\includegraphics[width=5.5in]{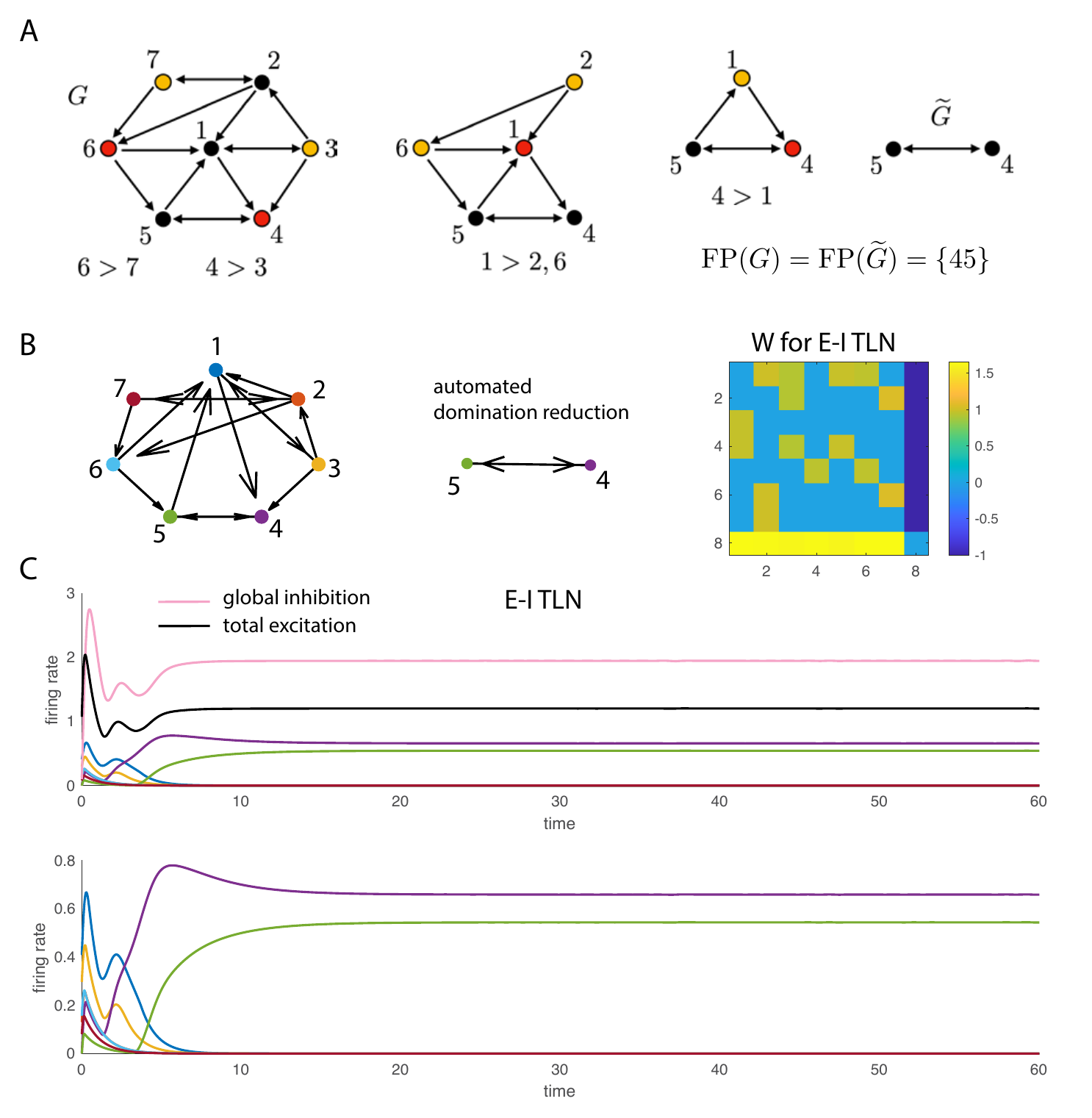}
\end{center}
\caption{{ E-I domination example 1.}}
\label{fig:ex1}
\end{figure}

\begin{figure}
\begin{center}
\includegraphics[width=5.5in]{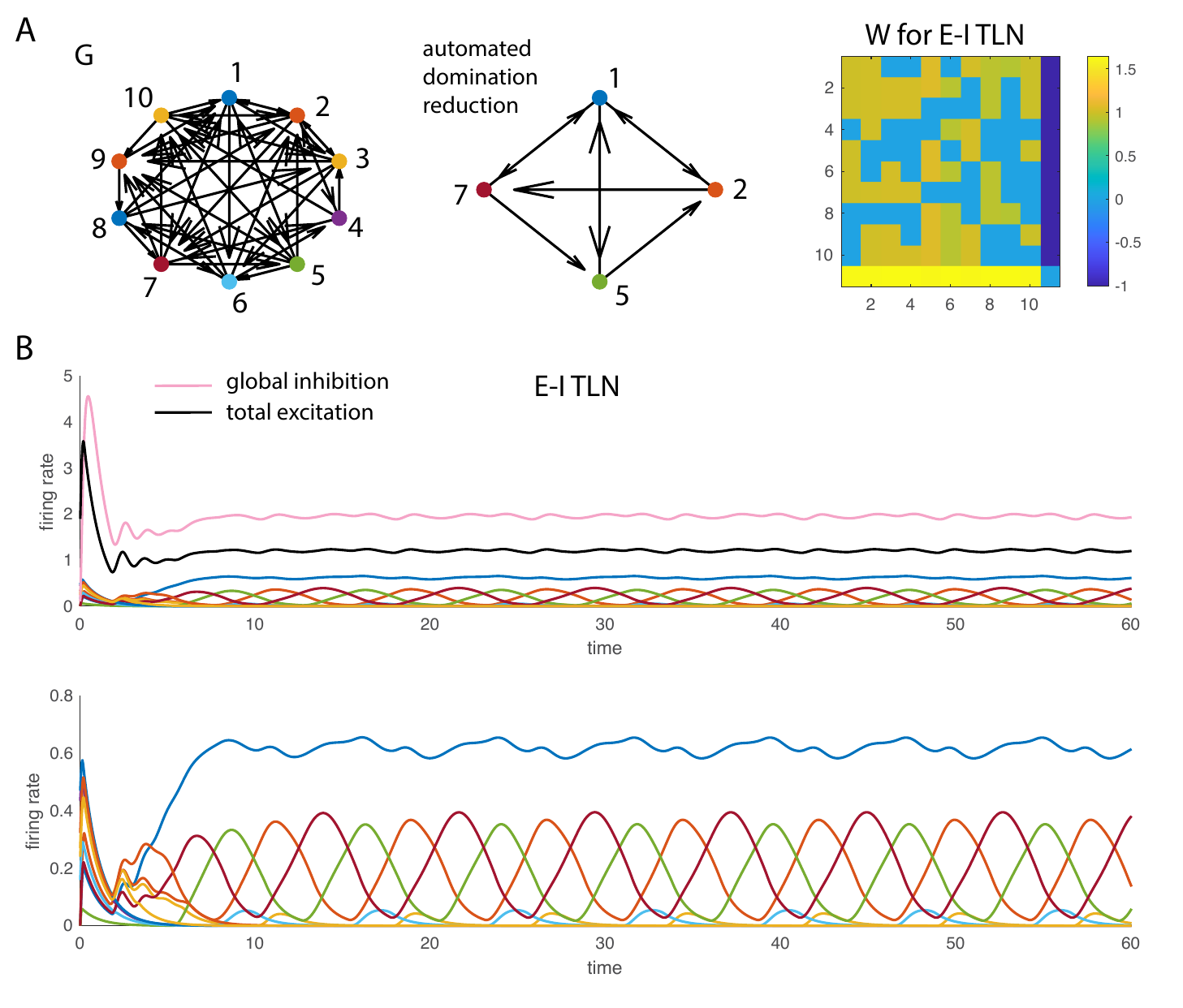}
\end{center}
\caption{{ E-I domination example 2.}}
\label{fig:ex2}
\end{figure}

As mentioned in the Introduction, our new domination theorems, Theorem 1 and Theorem 2, apply equally well to E-I TLNs. Figures~\ref{fig:ex1} and~\ref{fig:ex2} illustrate how well the behavior of an E-I network is predicted by the reduction. 

\subsection{Mapping E-I TLNs to gCTLNs} 
In the case where inhibition is significantly faster than excitation, so that $\tau_I \ll 1$, a separation of timescales argument shows that an E-I TLN precisely reduces to a gCTLN for the same graph $G$. 

For example, going back to the $n=3$ graph with edges $1 \leftrightarrow 2 \to 3 \to 1$, we obtain the reduction from an E-I network with weight matrix $W'$ to a gCTLN with weight matrix $W$:
$$W' = \left(\begin{array}{cccc} 
c_1 & a_2 & a_3 & -1 \\
a_1 & c_2 & 0 & -1 \\
0 & a_2 & c_3 & -1 \\
c_1 & c_2 & c_3 & 0 \end{array}\right)
\;\; \longmapsto\;\;
W = \left(\begin{array}{ccc} 
0 & a_2-c_2 & a_3-c_3  \\
a_1-c_1 & 0 & -c_3 \\
-c_1 & a_2-c_2 & 0  \\
\end{array}\right).
$$

The mapping between the parameters is as follows: if the E-I TLN has graph $G$ and parameters  $\{a_j,c_j,\theta,\tau_I\}$, then the corresponding gCTLN has the same graph $G$ and parameters $\{\varepsilon_j,\delta_j,\theta\}$, with:
\begin{eqnarray*}\label{eq:par-map}
\varepsilon_j &=& 1 + a_j - c_j,\\
\delta_j &=& c_j -1.
\end{eqnarray*}
Note that in order to ensure that $\varepsilon_j, \delta_j > 0$ in the gCTLN, we must have $a_j>0$ and $1 < c_j < 1+a_j$ in the E-I TLN. We also assume $b_I = 0$, as specified in the definition of the E-I TLN.

Conversely, given a gCTLN with graph $G$ and parameters $\{\varepsilon_j,\delta_j,\theta\}$, we can build an equivalent E-I TLN by choosing the same graph $G$ and parameters:
\begin{eqnarray*}\label{eq:par-map}
a_j &=& \varepsilon_j + \delta_j,\\
c_j &=& 1 + \delta_j.
\end{eqnarray*}

\begin{theorem}[Theorem 3]\label{thm1}
Let $(W,b)$ be a gCTLN with graph $G$ and parameters $\{\varepsilon_j,\delta_j,\theta\}$, and let $(W',b')$ be the corresponding E-I TLN under the above mapping, with graph $G$ and parameters $\{a_j,c_j,\theta,\tau_I\}$. Then $(W,b)$ and $(W',b')$ have the same fixed points in the following sense. There is a bijection, $\varphi: \operatorname{fixpts}(W,b) \to \operatorname{fixpts}(W',b')$, that sends
$$x^* = (x_1^*,\ldots,x_n^*) \mapsto \hat{x}^* = (x_1^*,\ldots,x_n^*,x_I^*),$$
where $x_I^* = \sum_{j=1}^n W'_{Ij}x_j^*$.
\end{theorem}

In other words, the fixed points of both networks exactly match on the excitatory neurons, $x_i$, for $1,\ldots,n$, and the inhibitory node $x_I$ for the E-I network has the unique value consistent with the excitatory neuron values $x_i^*$ at the fixed point.

\begin{proof}
Suppose $\hat{x}^* = (x_1^*,\ldots,x_n^*,x_I^*)$ is a fixed point of $(W',b')$. Then we must have $dx_I/dt = 0$, and so
$x_I^* = \sum_{j=1}^n W'_{Ij}x_j^*$ (since $b_I = 0$). Plugging this value of $x_I^*$ into the equations for $dx_i/dt = 0$, at the fixed point, we obtain:
\begin{eqnarray*}
0 &=& -x_i^* + \left[ \sum_{j=1}^n W'_{ij} x_j^* + W'_{iI}( \sum_{j=1}^n W'_{Ij}x_j^*) + \theta \right]_+, \; i=1,\ldots,n, \\
&=& -x_i^* + \left[ \sum_{j=1}^n (W'_{ij} + W'_{iI}W'_{Ij}) x_j^* + \theta \right]_+,\\
&=& -x_i^* + \left[ \sum_{j=1}^n W_{ij} x_j^* + \theta \right]_+.
\end{eqnarray*}
We can thus see that any fixed point of an E-I TLN $(W',b')$ corresponds to a fixed point of a gCTLN with $(W,b)$, where $b_i = b'_i = \theta$ for $i = 1,\ldots,n$, and 
$$W_{ij} = W'_{ij} + W'_{iI}W'_{Ij} = \left\{\begin{array}{cc} a_j - c_j & \text{ if } j \to i,\\
-c_j &  \text{ if } j \not\to i, \\
0 & \text{ if } i = j.\end{array}\right.$$
Now, using the mapping $a_j = \varepsilon_j + \delta_j$ and $c_j = 1+\delta_j$, we recognize that this
is precisely the $W$ matrix for the gCTLN with the same graph $G$ and parameters $\{\varepsilon_j,\delta_j,\theta\}$.

Conversely, starting with a fixed point $x^*$ of a gCTLN, it is easy to see that the augmented $(n+1)$-dimensional vector $\hat{x}^*$, as given by the map $\varphi$, is a fixed point of any E-I TLN with the same graph and parameters 
$\{a_j,c_j,\theta,\tau_I\}$ with $a_j = \varepsilon_j + \delta_j$ and $c_j = 1+\delta_j$. Note that the value of $\tau_I$ does not affect the mapping between the fixed points (though it may affect their stability).
\end{proof}

\subsection{Reduction of E-I TLNs to gCTLNs using fast-slow dynamics}
If $\tau_I \ll \tau_E =1$, we have a separation of timescales. Assuming the excitatory firing rates $x_1,\ldots,x_n$ change slowly compared to the inhibitory node $x_I$,  we can approximate the system~\eqref{eq:E-I-network} by assuming $x_I$ converges quickly to its steady state value,
$$x_I =  \left[ \sum_{j=1}^n W_{Ij} x_j \right]_+.$$
Furthermore, since $W_{Ij} > 0$ and $x_j \geq 0$, we can drop the nonlinearity to obtain simply:
$$x_I =  \sum_{j=1}^n W_{Ij} x_j,$$
even if we are not at a fixed point.
This inhibitory ``steady state," of course, depends on the dynamic variables $x_j$, so it will be continuously updated on the slower timescale that governs the excitatory dynamics.

We can now use the algebraic solution for $x_I$ and plug it into the $dx_i/dt$ equations, effectively reducing the system to only the first $n$ (excitatory) neurons. This yields,
$$\dfrac{dx_i}{dt} = -x_i + \left[ \sum_{j=1}^n \Wtil_{ij} x_j + b_i \right]_+,$$
where,
\begin{eqnarray}\label{eq:Wtil-mtx}
\Wtil_{ij} = \left\{\begin{array}{cc} W_{ij} + W_{iI}W_{Ij}  & \text{ if } i \neq j,\\
0 & \text{ if } i = j,\end{array}\right.
\end{eqnarray}
just as in the proof of Theorem 3, only now we do not require that $x$ is a fixed point. We thus see that, for fast enough $\tau_I$, we can expect the E-I TLN dynamics to effectively reduce to those of the corresponding gCTLN with matching fixed points.

\paragraph{How small does $\tau_I$ need to be for this to work?}

\begin{figure}
\begin{center}
\includegraphics[width=6in]{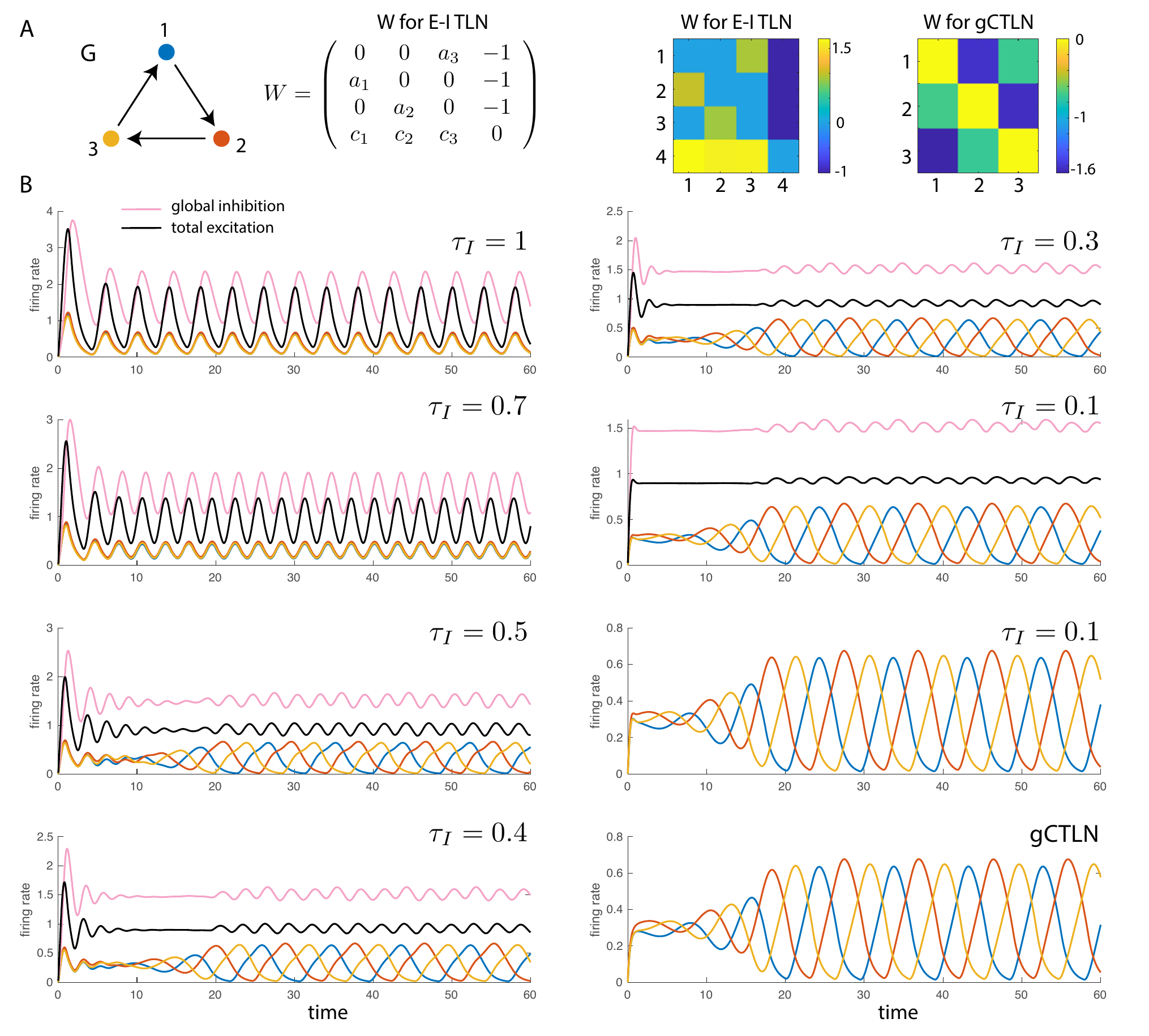}
\end{center}
\caption{{3-cycle E-I TLNs for a range of inhibitory timescales.}}
\label{fig:3-cycle}
\end{figure}

\begin{figure}
\begin{center}
\includegraphics[width=6in]{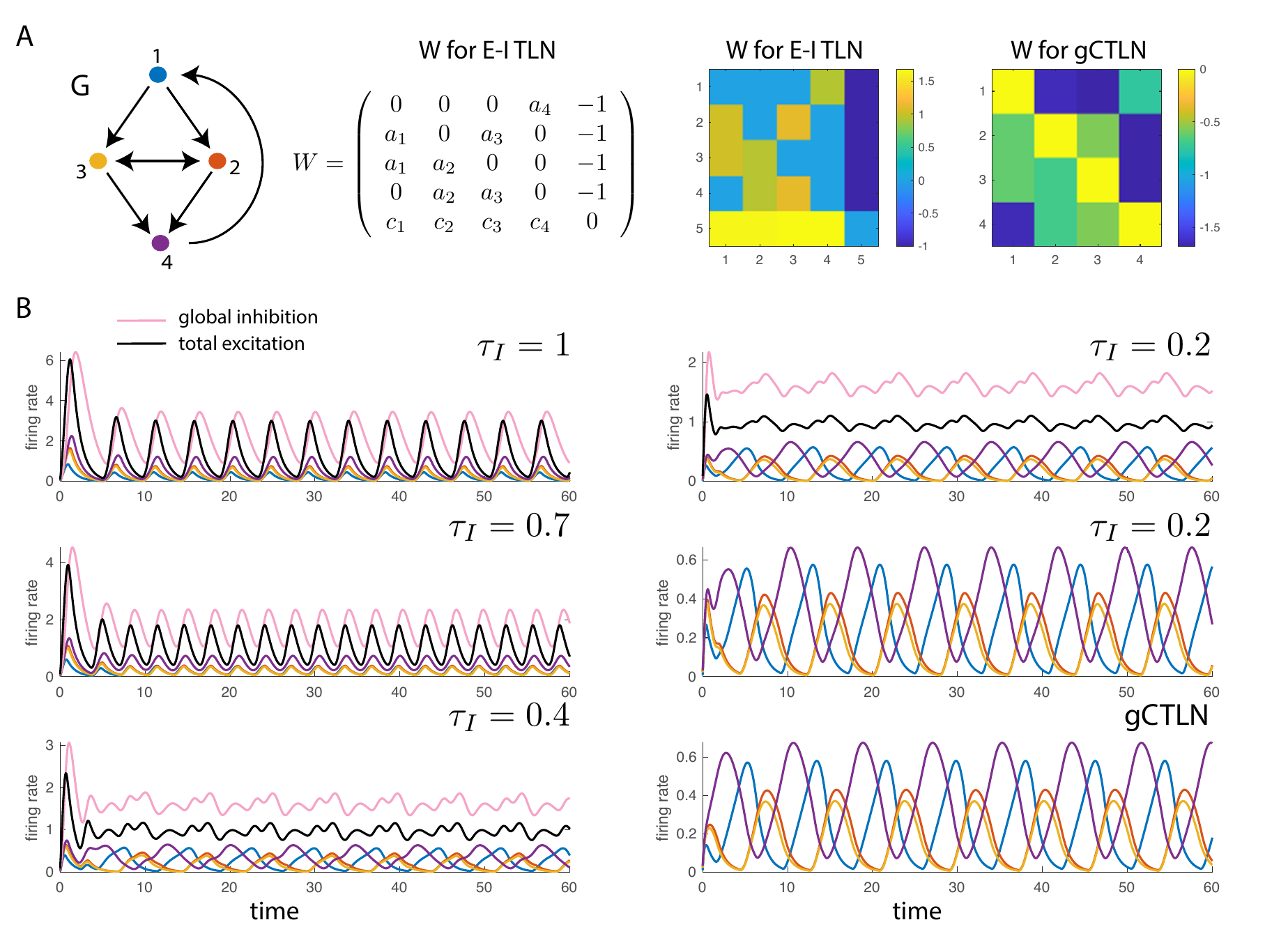}
\end{center}
\caption{{4-cycu E-I TLNs for a range of inhibitory timescales.}}
\label{fig:4-cycu}
\end{figure}

\begin{figure}
\begin{center}
\includegraphics[width=6in]{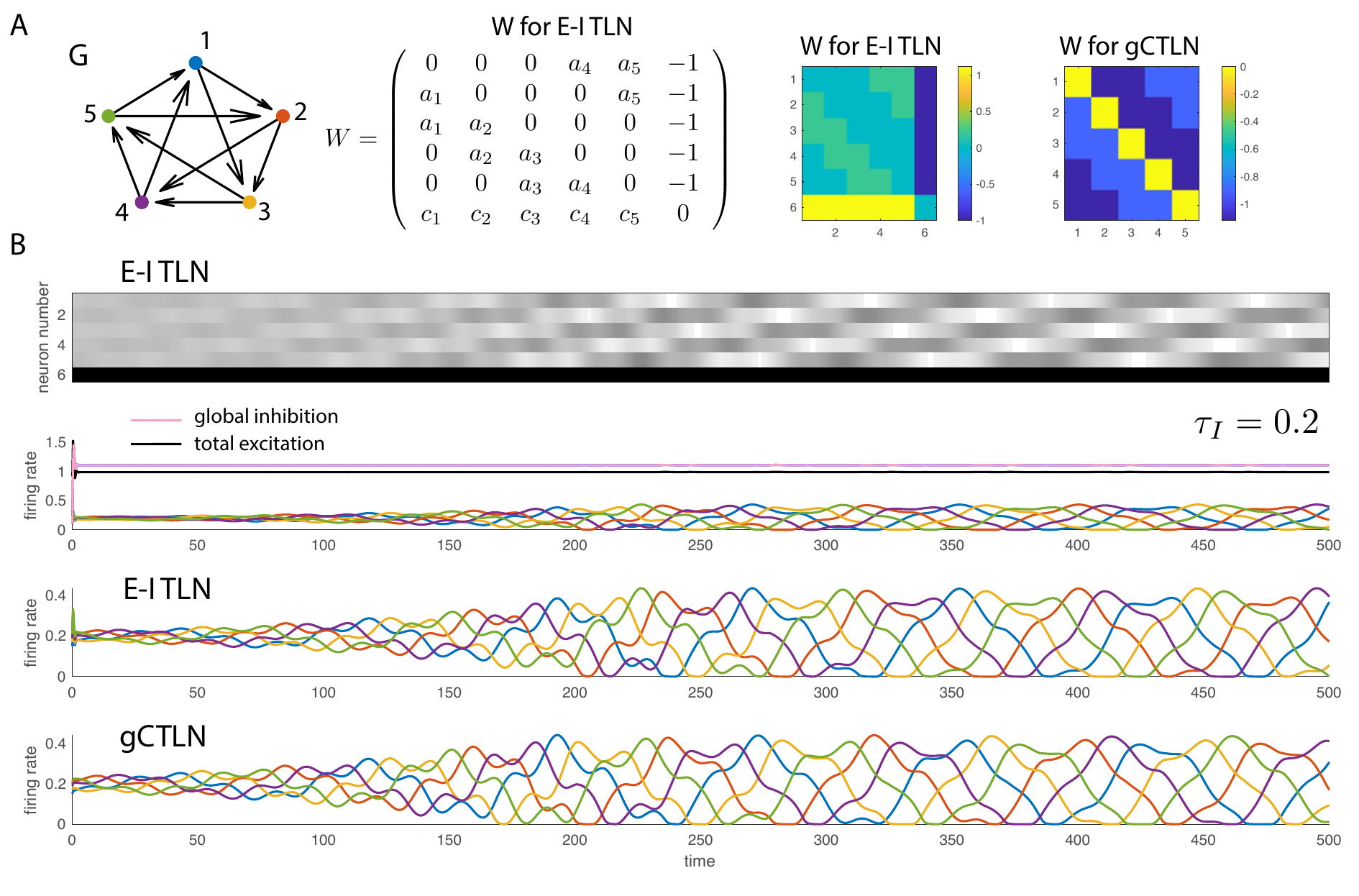}
\end{center}
\caption{{Gaudi E-I TLNs for a range of inhibitory timescales.}}
\label{fig:gaudi}
\end{figure}

\begin{figure}
\begin{center}
\includegraphics[width=6in]{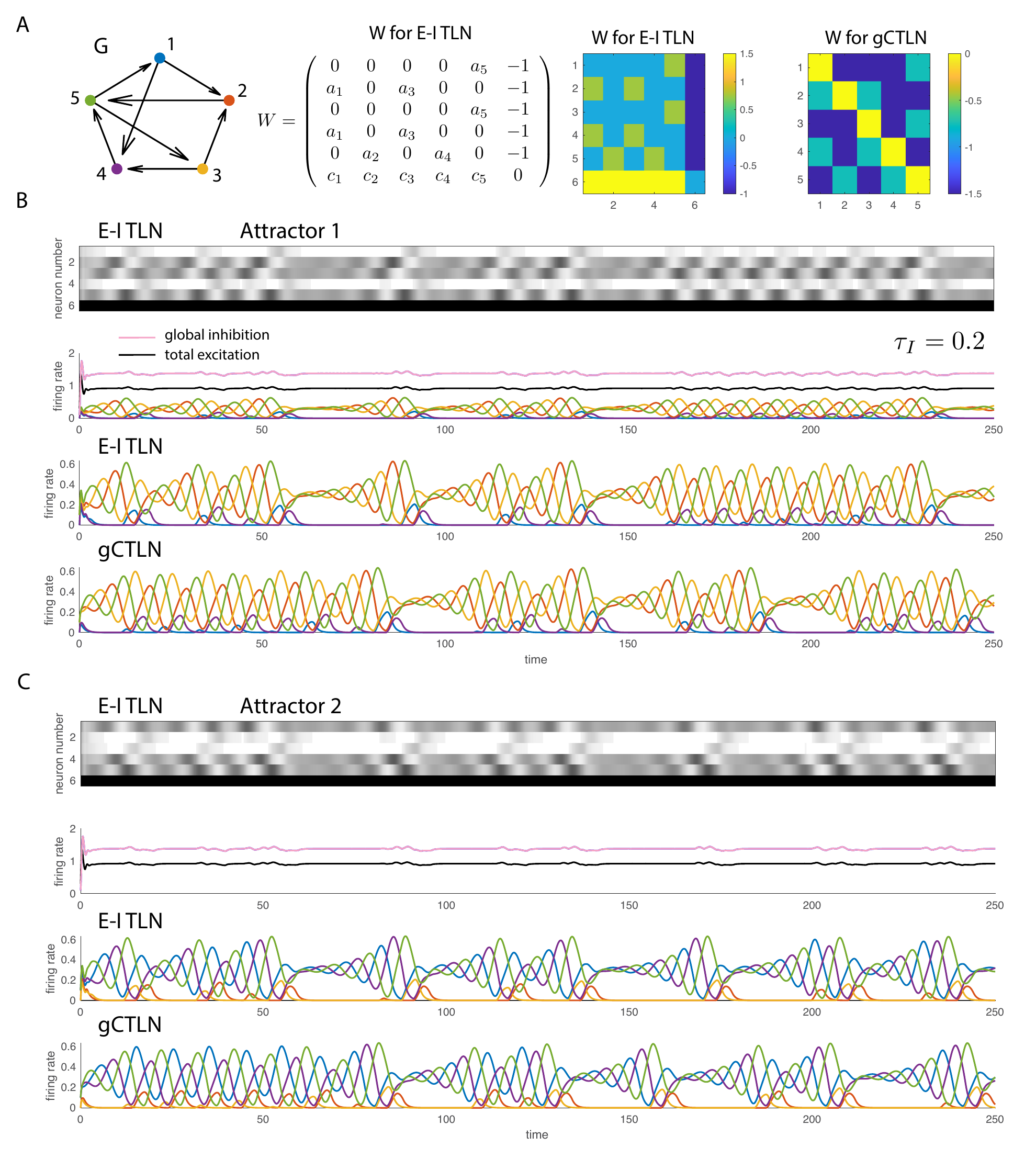}
\end{center}
\caption{{Baby chaos E-I TLNs for a range of inhibitory timescales.}}
\label{fig:baby-chaos}
\end{figure}

\begin{figure}[!h]
\begin{center}
\includegraphics[width=6in]{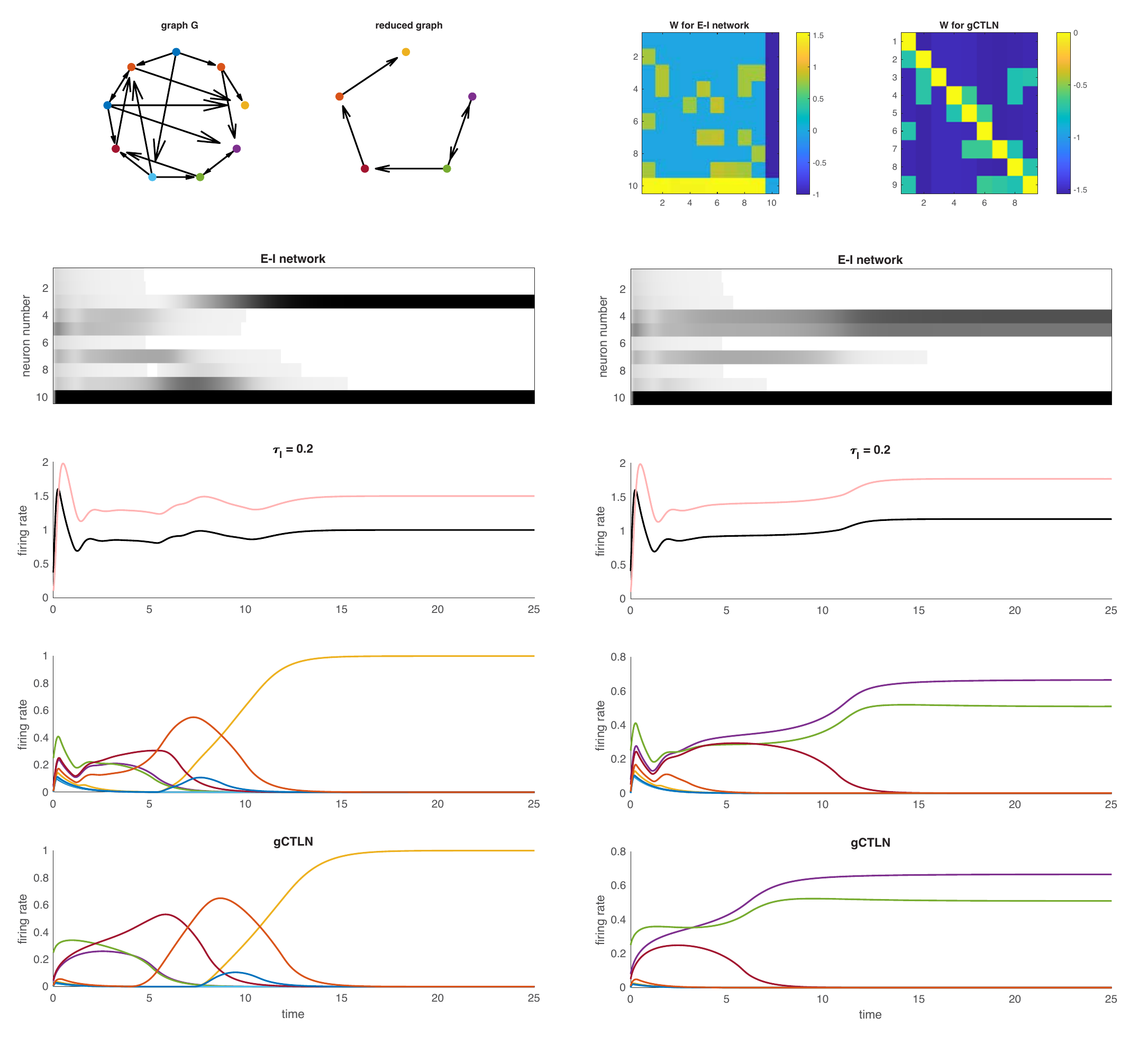}
\end{center}
\caption{Sample trajectories for E-I and gCTLN networks with the graph $G$ from Figure~\ref{fig:net+domination}D.}
\label{fig:dom-reduction}
\end{figure}

Figures~\ref{fig:3-cycle}-\ref{fig:baby-chaos} show example E-I TLNs for graphs corresponding to a 3-cycle, a 4-cycu, the ``Gaudi'' attractor, and ``baby chaos'' networks, which have previously been described for CTLNs in \cite{CTLN-diversity,Notices,extended-notices}, while Figure~\ref{fig:dom-reduction} shows the networks with graph $G$ given in Figure~\ref{fig:net+domination}D (with two initial conditions, one converging to each attractor). 
We can see in these examples that for $\tau_I = 1$, the same timescale as for excitation, the networks fall into E-I oscillations were all excitatory nodes are synchronized and the underlying graph structure is not reflected in the dynamics. However, as the timescale of inhibition gets faster, the dynamics are more and more similar to that of the corresponding gCTLN (last panel in each figure). Indeed, $\tau_I = 0.2$ appears to be fast enough in all of these cases. 

Note that the $W$ matrices displayed here for E-I TLNs have $0$ on the diagonal, as in the alternative convention with no self-excitation but a more complex inhibitory interaction term. This is equivalent to setting $W_{ii} = c_i$ on the diagonal.

%The $3$-cycle graph, $1 \to 2 \to 3 \to 1$, yields the following $4 \times 4$ weight matrix, where the index $4$ corresponds to the inhibitory node.
%$$W = \left(\begin{array}{cccc} 
%0 & 0 & a_3 & -1 \\
%a_1 & 0 & 0 & -1 \\
%0 & a_2 & 0 & -1 \\
%c_1 & c_2 & c_3 & 0 \end{array}\right)$$
%
%
%Simulation parameters for 3-cycle and 4-cycu: $e = .25+.2*\text{rand}(1,n)$
%$d = .6+.1*\text{rand}(1,n)$
%$X0 = .05*\text{rand}(n,1)$
%$b_i = 1, b_I = 0$
%-- Need to translate to $a_i$s and $c_i$s.
 
\pagebreak

%\carina{Shall we also conjecture that if $\tilG$ is a clique, then the corresponding fixed point is a global attractor?}

%\section{E-I TLNs}

%We consider recurrent networks with dynamics on the firing rates $x_1,\ldots,x_n$ of $n$ excitatory neurons with synaptic weights $W_{ij}$. There are two versions of the model: with and without an explicit inhibitory node, $x_I = x_{n+1}$. A specific network is defined by a pair $(W,b)$, where $W$ is a connectivity matrix and $b$ is a vector of external inputs or thresholds. \carina{Both models are trying to capture the situation in Figure 1A.}
%
%In both models, we restrict attention to threshold-linear networks built from a graph $G$. This means that the $W$ matrix is prescribed by $G$ and a set of weight parameters. In what follows, $G$ is a simple directed graph with vertex set $[n] = \{1,\ldots,n\}$. ({\it Simple} means that there are no multi-edges and no self loops.)

%\subsection{Two equivalent TLN models}

%\begin{figure}[!h]
%\begin{center}
%\includegraphics[width=5.5in]{figures/Fig1_29may2025.pdf}
%\end{center}
%\caption{{\bf E-I networks modeled with global inhibition.}}
%\label{fig:network-only}
%\end{figure}
%

\section{Proofs of domination theorems}\label{sec:dom-proofs}

\subsection{Fixed points of general TLNs}

Recall that a general TLN $(W,b)$ on $n$ neurons has dynamics,
\begin{eqnarray}\label{eq:general-TLN}
\tau_i\dfrac{dx_i}{dt} = -x_i + \left[ \sum_{j=1}^n W_{ij} x_j + b_i \right]_+,
\end{eqnarray}
where $x_i(t)$ is the activity of neuron $i$ at time $t$, $W$ is a real $n \times n$ connectivity matrix, $b \in \RR^n$ is a set of external inputs, and $\tau_i>0$ is the timescale of each neuron.

For fixed $W$ and $b$, we capture the fixed points of the dynamics via the set of all fixed point supports:
$$\FP(W,b) \od \{\sigma \subseteq [n] \mid \sigma =  \supp(x^*) \text{ for some fixed point } x^* \text{ of the TLN } (W,b) \},$$
where $\supp(x^*) = \{ i \mid x_i^* > 0\}$ and $[n] \od \{1,\ldots,n\}$.
For $\sigma \in \FP(W,b)$, the corresponding fixed point (for a {\it nondegenerate}\footnote{Namely, we need $\det(I-W_\sigma) \neq 0$ for each $\sigma \subseteq [n]$.} TLN) is easily recovered: 
$$x_\sigma^*  = (I-W_\sigma)^{-1}b_\sigma,$$ 
where $x_\sigma^*$ are the entries in the support $\sigma$, and $x_k^* = 0$ for all $k \notin \sigma$.
In particular, there is at most one fixed point per support.
Notice that the timescales $\tau_i$ do not affect the existence or values of the fixed points, which is why we don't include these parameters in the notation for $\FP(W,b)$. (They do, however, affect the stability and general behavior near the fixed points.)

The equations~\eqref{eq:general-TLN} can be rewritten as:
$$\tau_i \dfrac{dx_i}{dt} = -x_i + [y_i(x)]_+,$$
where
\begin{equation}\label{eq:y_i}
y_i(x) = \sum_{\ell=1}^n W_{i\ell}x_\ell + b_i.
\end{equation}
Clearly, if $x^*$ is a fixed point of $(W,b)$, then $x^*_i = [y^*_i]_+$ for all $i \in [n]$, where $y_i^* = y_i(x^*)$. 

With this notation, we have the following lemma:
\begin{lemma}
Let $x^* \in \RR^n_{\geq 0}$ have support $\sigma \subseteq [n]$. Then $x^*$ is
a fixed point of $(W,b)$ if and only if 
\begin{itemize}
\item[(i)] $x_i^* = y_i^* > 0$ for all $i \in \sigma$ (on-neuron conditions), and
\item[(ii)] $y_k^* \leq 0$ for all $k \not\in \sigma$ (off-neuron conditions).
\end{itemize}
\end{lemma}

\noindent This simple characterization of fixed points will allow us to rule in and rule out fixed points with various supports in the presence of domination.

\subsection{Input domination for general TLNs $(W,b)$}
All results in this section hold for general TLNs of the form~\eqref{eq:general-TLN} . In particular, there is no requirement that the timescales $\tau_i > 0$ are the same, nor that the self-excitation $W_{ii} = 0$. All weights $W_{ij}$ can in principle be positive, negative, or zero. We do require nondegeneracy of the TLN for some of the results about fixed points.

For a general TLN, we define ``input domination'' as follows:
\begin{definition}
Let $(W,b)$ be a TLN on $n$ nodes. We say that $k$ {\em input dominates} $j$ if the following three conditions hold:
\begin{itemize}
\item[(i)] $W_{ki} \geq W_{ji}$ for each $i \in [n]\setminus\{j,k\}$,
\item[(ii)] $W_{kj} > -1 +W_{jj}$,
\item[(iii)] $W_{jk} < -1 + W_{kk}$, and
\item[(iv)] $b_k \geq b_j$.
\end{itemize}
In the case where $W$ has zero diagonal, (ii)-(iii) become $W_{kj} > -1 > W_{jk}.$
\end{definition}

The idea here is that node $k$ input dominates $j$ if $k$ receives both more recurrent input (as in (i)) and more external input (as in (iv)) than $j$ does. Additionally, it is important that the $j \to k$ weight, $W_{kj}$, is greater than $j$'s self inhibition, $-1+W_{jj}$ (as in (ii)); while for the $k \to j$ weight, $W_{jk}$, it is the other way around (as in (iii)).

With this definition, we have the following key lemma:

\begin{lemma}\label{lemma:input-domination}
Let $(W,b)$ be a TLN, and suppose $k$ input dominates $j$. Then,
$$ -x_j + y_j \leq -x_k + y_k \;\; \text{ for all } \;\; x \in \RR_{\geq 0}^n.$$
Moreover, if either $x_j >0$, $x_k>0$, or $b_k > b_j$, then the above inequality is strict
and 
$$ -x_j + y_j < -x_k + y_k \;\; \text{ for all } \;\; x \in \RR_{\geq 0}^n.$$
Finally, if $0 \leq x_j \leq x_k$, then we have
$y_j \leq y_k.$
\end{lemma}
Note that, for $x \in \RR^n_{\geq 0}$, the condition $0 \leq x_j \leq x_k$ is automatically satisfied if $x_j = 0$.

\begin{proof}
First, observe that:
\begin{eqnarray*}
y_j &=& \sum_{i \neq j,k} W_{ji} x_i + W_{jj} x_j + W_{jk} x_k + b_j,\\
y_k &=& \sum_{i \neq j,k} W_{ki} x_i + W_{kj} x_j + W_{kk} x_k + b_k.
\end{eqnarray*}
This implies:
\begin{eqnarray*}
y_j + x_k &=& \sum_{i \neq j,k} W_{ji} x_i + W_{jj} x_j + (1 + W_{jk}) x_k + b_j,\\
y_k + x_j &=& \sum_{i \neq j,k} W_{ki} x_i + (1 + W_{kj}) x_j + W_{kk} x_k + b_k.
\end{eqnarray*}
Now, since $k$ input dominates $j$, then for all $x \in  \RR_{\geq 0}^n$ (which means all $x_i$ are nonnegative), we have the following four inequalities:
$$\sum_{i \neq j,k} W_{ji} x_i  \leq  \sum_{i \neq j,k} W_{ki} x_i, \;\;\; b_j \leq b_k,$$
and
$$W_{jj} x_j \leq (1 + W_{kj}) x_j, \;\; \; (1 + W_{jk}) x_k \leq W_{kk} x_k.$$
Note that although conditions (ii)-(iii) of input domination give strict inequalities $W_{jj} < 1+W_{kj}$ and $1+W_{jk} < W_{kk}$, once we multiply by $x_j$ and $x_k$ the inequalities become non-strict, as we could have $x_j = 0$ or $x_k = 0$. As a result of all four inequalities, we can conclude that
$y_j + x_k \leq y_k + x_j$, and thus
$$-x_j + y_j \leq -x_k + y_k,$$
as desired. Moreover, if any of the four inequalities is strict, we have $y_j + x_k < y_k + x_j$ and thus 
$-x_j + y_j < -x_k + y_k.$ This occurs if either $x_j > 0$ or $x_k > 0$ or $b_k > b_j$ (only one of the three has to hold).
Finally, if $x_j \leq x_k$, then 
$$y_j \leq x_j-x_k + y_k \leq y_k$$ 
for all $x \in  \RR_{\geq 0}^n$ .
\end{proof}

As an immediate consequence of Lemma~\ref{lemma:input-domination}, we have the following:

\begin{corollary}\label{cor:input-domination}
Suppose $k$ input dominates $j$ in the TLN $(W,b)$. Then there can be no fixed point $x^*$ of $(W,b)$ with $x_j^* > 0$.
\end{corollary}

\begin{proof}
Suppose $x^*$ is a fixed point of $(W,b)$ with $x_j^*>0$. Then $x_\ell^* = [y_\ell^*]_+$ for all $\ell \in [n]$. In particular, 
$x_j^* = [y_j^*]_+ = y_j^* > 0$ and $x_k^* = [y_k^*]_+.$ By Lemma~\ref{lemma:input-domination}, we have
$$0 = -x_j^* + y_j^* < -x_k^* + y_k^* \leq -x_k^* + [y_k^*]_+ = 0.$$
The above equation thus gives $0 < 0$, a contradiction. We conclude that we must have $x_j^* = 0$ at every fixed point of $(W,b)$.
\end{proof}

Note in the above proof that if $x_j^* = 0$ at the fixed point, then $y_j^* \leq 0$ and we would not be able to conclude that $-x_j^* + y_j^* = 0$, which was the source of the contradiction.

Corollary~\ref{cor:input-domination} tells us that if $j$ is a dominated node, then
$$\FP(W,b) \subseteq \FP(W|_{[n] \setminus j}, b|_{[n] \setminus j}),$$
because $j$ does not participate in any fixed points. However, it could be that there are fixed points in the reduced network $(W|_{[n] \setminus j}, b|_{[n] \setminus j})$ that do not survive to $(W,b)$, because the presence of node $j$ kills them. Our first domination theorem assures us that this is not the case, and therefore $j$ can be removed from the network without altering the set of fixed points.

\begin{theorem}\label{thm:input-domination}
Let $(W,b)$ be a nondegenerate TLN on $n$ nodes, and suppose $k$ input dominates $j$. Then,
$$\FP(W,b) = \FP(W|_{[n] \setminus j}, b|_{[n] \setminus j}).$$
\end{theorem}

\begin{proof}
By Corollary~\ref{cor:input-domination} we have $\FP(W,b) \subseteq \FP(W|_{[n] \setminus j}, b|_{[n] \setminus j}).$ To see the reverse inclusion, let $\sigma \in \FP(W|_{[n] \setminus j}, b|_{[n] \setminus j})$ and consider a fixed point $x^*$ of $(W|_{[n] \setminus j}, b|_{[n] \setminus j})$ with support $\supp(x^*) = \sigma \subseteq [n]\setminus j$. To see whether the fixed point ``survives'' in the larger network, so that $\sigma \in \FP(W,b)$, it suffices to verify that $y_j(\hat x^*) \leq 0$, where $\hat x^*$ is the augmented vector in $\RR^n$ obtained by setting $\hat x_j^*=0$, and $\hat x_i^* = x_i^*$ for all $i \in [n]\setminus j$. (I.e., we need only check that the off-neuron condition holds for the added node $j$, so that $j$ does not get activated by activity at the fixed point -- which would contradict the existence of the fixed point in $(W,b)$.)

Since $k$ input dominates $j$, Lemma~\ref{lemma:input-domination} tells us that for all $x \in \RR^n_{\geq 0}$ for which $x_j = 0$, we have $y_j \leq -x_k + y_k$. Therefore, since $\hat x_j^* = 0$, we know that this holds for $\hat x^*$, and so:
$$\hat y_j^* \leq -\hat x_k^* + \hat y_k^*,$$
where $\hat y_j^* = y_j(\hat x^*)$ and $\hat y_k^* = y_k(\hat x^*)$. 
Moreover, since $x^*$ is a fixed point of $(W|_{[n] \setminus j}, b|_{[n] \setminus j})$, we know that $x_k^* = [y_k^*]_+$ and thus $\hat x_k^* = [\hat y_k^*]_+$. Recalling that $\hat y_k^* \leq [\hat y_k^*]_+$ (by definition of the ReLU nonlinearity), it follows that:
$$\hat y_j^* \leq -\hat x_k^* + [\hat y_k^*]_+ = 0.$$ 
Thus, $\hat y_j^* \leq 0$, as desired, and we can conclude that 
$\hat x^*$ is a fixed point of $(W,b)$ with support $\sigma \in \FP(W,b)$.
\end{proof}

As the above proof makes clear, it is not only that the networks $(W,b)$ and $(W|_{[n] \setminus j}, b|_{[n] \setminus j})$ have the same fixed point {\it supports}. The actual values of the fixed points of the larger network are identical to those of the smaller network, except for the added entry $\hat x_j^* = 0.$

\subsection{Application to graph-based networks}\label{sec:thm1-proof}

For graph-based networks, such as gCTLNs and E-I TLNs, we have a notion of graphical domination that is defined on the underlying directed graph $G$. In this setting, graphical domination in $G$ implies input domination in the associated TLN.

\begin{lemma} \label{lemma:input-to-graphical}
Consider a directed graph $G$ on $n$ nodes. Suppose 
 $k$ graphically dominates $j$ for some $j,k \in [n]$. Then, for any gCTLN or E-I TLN with graph $G$, $k$ input dominates $j$.
 \end{lemma}
 
 \begin{proof}
 We will prove this first for gCTLNs, then for E-I TLNs.
 
 Suppose $(W,b)$ is the TLN corresponding to a gCTLN with parameters $\{\varepsilon_i,\delta_i, \theta\}$, with $\varepsilon_i,\delta_i>0$ for all $i \in [n]$. This means that for any $i,j \in [n]$, $W_{ij} = -1+\varepsilon_j$ when $j \to i$ in $G$, $W_{ij} = -1-\delta_j$ when $j \not\to i$, and $W_{ii} = 0$ .
 If $k$ graphically dominates $j$, then we obtain:
 \begin{eqnarray*}
 &\text{(i)}& W_{ki} \geq W_{ji} \text{ for  each } i \in [n] \setminus \{j,k\} \;\; (\text{since } i \to j \Rightarrow i \to k)\\
 &\text{(ii)}& W_{kj} > -1 \;\; (\text{since } j \to k),\\
 &\text{(iii)}& W_{jk} < -1 \;\; (\text{since } k \not\to j),\\
 &\text{(iv)}& b_k \geq b_j \;\; (\text{since } b_k = b_j = \theta).
 \end{eqnarray*}
Recalling that $W_{jj} = W_{kk} = 0$, we see that the conditions for input domination are precisely satisfied, so  that $k$ input dominates $j$.

To see the result for E-I TLNs, let $(W,b)$ be the E-I TLN corresponding to graph $G$ with parameters $\{a_i, c_i, \theta\}$. Since $G$ has $n$ nodes, $W$ is an $(n+1) \times (n+1)$ matrix and $b \in \RR^{n+1}$, with position $n+1$ corresponding to the inhibitory ``$I$'' node.
We thus have $W_{ij} = a_j$ when $j \to i$ in $G$, $W_{ij} = 0$ when $j \not\to i$, $W_{iI} = -1$, $W_{Ij} = c_j$, $W_{ii} = -W_{iI}W_{Ii} = c_i$, $b_i = \theta$ for all $i \in [n]$, and $b_I = 0$. Furthermore, recall that for E-I TLNs we require $a_j > 0$ and $1 < c_j < 1+a_j$ for all $j \in [n]$ (this is equivalent to the requirement that $\varepsilon_j, \delta_j > 0$ in the gCTLN).

Now we can check each of the four conditions input domination. Since $W_{ki} = a_i$ or $0$, and $W_{ji} = a_i$ or $0$, we clearly satisfy condition (i) $W_{ki} \geq W_{ji}$ for all $i \neq j,k$, since $i \to j \Rightarrow i \to k$. Condition (iv) is also easy: it holds since $b_j = b_k = \theta$. This leaves conditions (ii) and (iii), where we now must remember that $W_{jj} = c_j$, $W_{kk} = c_k$, and $W_{kj} = a_j$ (since $j \to k$) while $W_{jk} = 0$ (since $k \not\to j$). It follows that,
$$\text{(ii)} \;\; W_{kj} = a_j > -1 + c_j = -1+W_{jj},$$
since we have required $c_j < 1+a_j$. Similarly,
$$\text{(iii)} \;\; W_{jk} = 0 < -1 + c_k = -1+W_{kk},$$
since we required $c_k > 1$. We again conclude that $k$ input dominates $j$.
 \end{proof}

Note that because input domination was defined independently of the timescales $\tau_i$, the results stemming from this property hold for E-I TLNs despite the fact that the inhibitory timescale, $\tau_I$, is distinct from that of the excitatory nodes. The reason the timescales don't matter is that the results are all about the set of fixed points of a TLN. This set is independent of the choice of timescales, although the stability of a fixed point does potentially depend on the $\tau_i$s.

We can now prove the following theorem about graphical domination in gCTLNs and E-I TLNs. Note that in both cases, the network corresponding to the reduced graph $G|_{[n]\setminus j}$ must be viewed as having the same set of parameters as the original network, restricted to the index set $[n] \setminus j$. Furthermore, the gCTLN parameters $\{\varepsilon_i,\delta_i, \theta\}$ must satisfy the usual constraints $\varepsilon_i,\delta_i>0$, and the E-I TLN parameters
$\{a_i, c_i, \theta\}$ must satisfy the equivalent constraints $a_j > 0$ and $1 < c_j < 1+a_j$, as these were needed in the proof of Lemma~\ref{lemma:input-to-graphical}.

\begin{theorem}[Theorem 1] \label{thm:domination-both}
Suppose $j$ is a dominated node in a directed graph $G$.
Then the fixed points of any gCTLN (or E-I TLN) constructed from $G$ satisfy 
$$\FP(G) = \FP(G|_{[n]\setminus j}),$$
for any choice of gCTLN parameters $\{\varepsilon_i, \delta_i, \theta\}$ 
(or $\{a_i,c_i,\theta\}$).
\end{theorem}

\begin{proof}
Let $(W,b)$ be the gCTLN obtained from $G$ with parameters $\{\varepsilon_i, \delta_i, \theta\}$.
Since $j$ is a dominated node, there exists $k \neq j$ such that $k$ graphically dominates $j$. By Lemma~\ref{lemma:input-to-graphical}, we know that $k$ input dominates $j$. Applying Theorem~\ref{thm:input-domination} to the TLN $(W,b)$, we see that
$$\FP(G) = \FP(W,b) = \FP(W|_{[n] \setminus j}, b|_{[n] \setminus j}).$$
The theorem now follows from observing that the network $(W|_{[n] \setminus j}, b|_{[n] \setminus j})$ is precisely the gCTLN for the restricted graph, $G|_{[n]\setminus j}$, with the same (restricted) set of parameters.

The same argument shows that the result holds for any E-I TLN obtained from $G$.
\end{proof}

\subsection{Uniqueness of the reduced graph $\tilG$}\label{sec:uniqueness}

Using Theorem 1, it is clear that for a gCTLN (or an E-I TLN) $\FP(G) = \FP(\tilG)$. Also, it is clear that $\tilG$ cannot be further reduced by removing dominated nodes, because it is domination free. But it is not at all clear whether or not $\tilG$ is unique! If we remove dominated nodes in a different order, might we end up with a different reduced graph $\tilG$?
Note that $\tilG$ could involve nodes that do not appear in $\FP(G)$, even if $\tilG$ is unique. Consider graph E1[4] from \cite{rule-of-thumb}.\footnote{The classification of oriented graphs for $n \leq 5$ can found in the Supporting Information, towards the end of the arXiv version.} This graph is domination free but has $\FP(G) = \{123\}$, with nodes $4$ and $5$ not appearing in $\FP(G)$ but also not dominated.

It turns out that $\tilG$ is indeed unique. A key component to proving uniqueness is showing that if a node $j$ is dominated by another node $k$ that gets removed, then whoever dominated $k$ also dominates $j$. So once a node is ``dominated'' within a graph $G$, it will continue to be dominated at further steps in the reduction process until it is removed. This is the content of Corollary~\ref{cor:domination}, below. 

We will prove this fact via two simple lemmas. The first lemma is on the transitivity of domination. The second is about its inheritance to smaller graphs. Note that everything in this section is strictly about the graph $G$ and its reduction $\tilG$. There is no need to consider an associated gCTLN. 

\begin{lemma}\label{lemma:dom-transitivity}
Suppose $\ell$ graphically dominates $k$ and $k$ graphically dominates $j$ with respect to $G$. Then $\ell$ graphically dominates $j$ with respect to $G$.
\end{lemma}

\begin{proof}
Since $j,k,\ell$ are all vertices of $G$, the graphical domination assumptions imply that $j \to k \to \ell$ and $\ell \not\to k \not\to j$. They also tell us that for each $i \in [n] \setminus \{k,\ell\}$, if $i \to k$ then $i \to \ell$; and for each $i \in [n]\setminus \{j,k\}$, if $i \to j$ then $i \to k$. From here we can conclude that $j \to \ell$, and if $i \in [n]\setminus \{j,k,\ell\}$, then $i \to j$ implies $i \to \ell$. Moreover, if $\ell \to j$ then we'd have $\ell \to k$, a contradiction; so we can also conclude that $\ell \not\to j$. It follows that $\ell$ graphically dominates $j$ with respect to $G$. 
\end{proof}

\begin{lemma}\label{lemma:dom-inheritance}
If $k$ graphically dominates $j$ with respect to $G$, then $k$ graphically dominates $j$ with respect to $G|_{\omega}$ for any $\omega \subseteq [n]$ that contains both $j$ and $k$.
\end{lemma}

\begin{proof}
This is a trivial consequence of the definition of domination with respect to a full graph (i.e., inside-in domination). By assumption, $j \to k$, $k \not\to j$, and for each $i \in [n]\setminus\{j,k\}$, if $i \to j$ then $i \to k$. Now, if $\omega \subseteq [n]$, and $j,k \in \omega$, then we still have $j \to k$, $k \not\to j$, and it's clear that for each $i \in \omega\setminus\{j,k\}$, if $i \to j$ then $i \to k$.
\end{proof}

Putting together these two lemmas, we get the following useful corollary:

\begin{corollary}\label{cor:domination}
If $j$ is a dominated node in $G$, and $G|_{[n] \setminus d}$ is obtained from $G$ by removing another dominated node $d \neq j$, then $j$ is also a dominated node in $G|_{[n] \setminus d}$ (even if $d$ dominated $j$ in $G$).
\end{corollary}

\begin{proof}
If the removed node $d$ dominates $j$ in $G$, then $d$ must be dominated by some other node $\ell$, and by Lemma~\ref{lemma:dom-transitivity} we have that $\ell$ dominates $j$. If, on the other hand, $j$ was not dominated by $d$, then it is dominated by another node $k$ in $G$ and by Lemma~\ref{lemma:dom-inheritance} we know that $k$ still dominates $j$ in $G|_{[n] \setminus d}$. Either way, $j$ continues to be a dominated node in the subgraph.
\end{proof}

We are now ready to prove uniqueness of the reduced graph $\tilG$, Theorem~\ref{thm:dom-uniqueness}. The key is showing that no matter what the order of removal, the same set of nodes gets removed before arriving to a domination-free graph.

%\begin{theorem}
%Let $G$ be a directed graph on $n$ nodes. Then the reduced graph $\tilG$ is unique. Moreover, $\FP(G) = \FP(\tilG)$.
%\end{theorem}

\begin{proof}[Proof of Theorem~\ref{thm:dom-uniqueness}] 
Let $\tilG$ be a reduced graph obtained from $G$ by removing nodes $j_1,\ldots,j_m$, in that order. In other words, there is a decreasing filtration of graphs,
$$G \supseteq G|_{[n]\setminus \{j_1\}} \supseteq G|_{[n]\setminus \{j_1,j_2\}} \supseteq \cdots
\supseteq G|_{[n]\setminus \{j_1,j_2,\ldots,j_m\}} = \tilG,$$
where $j_1$ is graphically dominated by some $k_1$ in $G$, $j_2$ is graphically dominated by some $k_2$ in $G|_{[n]\setminus \{j_1\}}$, and so on. The sequence stops at $\tilG$, as it is domination free.

Now suppose $\widetilde{H}$ is another reduced graph obtained from $G$ by removing dominated nodes $\ell_1,\ldots,\ell_p$, in that order. This time the filtration is,
$$G \supseteq G|_{[n]\setminus \{\ell_1\}} \supseteq G|_{[n]\setminus \{\ell_1,\ell_2\}} \supseteq \cdots
\supseteq G|_{[n]\setminus \{\ell_1,\ell_2,\ldots,\ell_p\}} = \widetilde{H}.$$
WLOG, suppose $p \leq m$. We will show that for each $i=1,\ldots,m$, $j_i \in \{\ell_1,\ell_2,\ldots,\ell_p\}.$ This in turn will imply that $p = m$ and $\{\ell_1,\ell_2,\ldots,\ell_p\} = \{j_1,j_2,\ldots,j_m\}$, hence $\widetilde{H} = \tilG.$

First, consider $j_1$. Since $j_1$ is dominated in $G$, by Corollary~\ref{cor:domination} it will remain dominated in $G|_{[n]\setminus \{\ell_1\}}, G|_{[n]\setminus \{\ell_1,\ell_2\}},$ and so on all the way to $G|_{[n]\setminus \{\ell_1,\ldots,\ell_p\}}$, unless $j_1 \in  \{\ell_1,\ldots,\ell_p\}.$ Since $\widetilde{H} = G|_{[n]\setminus \{\ell_1,\ldots,\ell_p\}}$ is domination free, we can conclude that $j_1 \in  \{\ell_1,\ldots,\ell_p\}.$ Say, $j_1 = \ell_q$.

Now consider $j_2$. Since $j_2$ is graphically dominated in $G|_{[n]\setminus \{j_1\}}$, and $j_1 = \ell_q$, by Lemma~\ref{lemma:dom-inheritance} we know that $j_2$ will also be dominated in $G|_{[n]\setminus \{\ell_1,\ldots,\ell_q\}}$, unless $j_2 \in \{\ell_1,\ldots,\ell_q\}.$ Furthermore, by Corollary~\ref{cor:domination}, $j_2$ will remain dominated in the subsequent graphs of the  filtration all the way down to $\widetilde{H} = G|_{[n]\setminus \{\ell_1,\ldots,\ell_p\}}$, unless $j_2 \in \{\ell_{q+1},\ldots,\ell_p\}.$ Since $\widetilde{H}$ is domination free, we can conclude that $j_2 \in \{\ell_1,\ldots,\ell_p\}.$
Continuing in this fashion, we can show that all of the nodes $j_1,\ldots,j_m$ are in the set $\{\ell_1,\ldots,\ell_p\}$, as desired.
%Finally, using the new domination theorem at each step of the filtration for $\tilG$, we easily obtain that $\FP(G) = \FP(\tilG)$.
\end{proof}

%\carina{Need to update and include this info for each simulation.}
%
%For each network, we show (a) the original graph $G$, (b) the reduced graph $\tilG$, (c) the fixed point supports $\FP(G)$ (which, because of Theorem~\ref{thm:dom-uniqueness}, are the same for $G$ and $\tilG$), and (d) the results of a simulation for a gCTLN on $G$. The gCTLN parameters were selected randomly, with
%$$\varepsilon_i = 0.25+0.2\eta, \;\; \delta_i = 0.6+0.05\xi,$$
%where $\eta, \xi$ are distributed uniformly at random in $[0,1]$. We also used $b_i = \theta = 1$ for all $i=1,\ldots,n$. 
%
%In the simulations, we have chosen an initial condition $x(0)$ where each entry, $x_i(0)$, is distributed uniformly at random in $[0,0.5]$. However, for the nodes that remain in the reduced graph, $\tilG$, we purposefully set $x_i(0) = 0$ so as to bias the initial conditions away from these nodes. (The only exception is in Example 3b, where we gave $x_8(0)$ a high value in order to converge to the second attractor for this network.) Despite the bias, we see in all cases that the activity converges to an attractor whose activity is concentrated on the subgraph corresponding to $\tilG$. In Examples 1-5, the activity converges to a stable fixed point attractor supported on a clique in $\tilG$, with Examples 3, 4, and 5 each having more than one such attractor. In Examples 6-9, the activity converges to a dynamic attractor (a limit cycle), with the most active neurons all belonging to the $\tilG$ subgraph.

\bibliographystyle{unsrt}
\bibliography{CTLN-refs-JLA}

\begin{thebibliography}{1}

\bibitem{fp-paper}
C.~Curto, J.~Geneson, and K.~Morrison.
\newblock Fixed points of competitive threshold-linear networks.
\newblock {\em Neural Comput.}, 31(1):94--155, 01 2019.

\bibitem{Notices}
C.~Curto and K.~Morrison.
\newblock Graph rules for recurrent neural network dynamics.
\newblock {\em Notices of the American Mathematical Society}, 70(04):536--551,
  2023.

\bibitem{extended-notices}
C.~Curto and K.~Morrison.
\newblock Graph rules for recurrent network dynamics: extended version, 2023.
\newblock arXiv preprint arXiv:2301.12638.

\bibitem{CTLN-diversity}
K.~Morrison, A.~Degeratu, V.~Itskov, and C.~Curto.
\newblock Diversity of emergent dynamics in competitive threshold-linear
  networks.
\newblock {\em SIAM J. Applied Dynamical Systems}, 2024.

\bibitem{rule-of-thumb}
C.~Parmelee, S.~Moore, K.~Morrison, and C.~Curto.
\newblock Core motifs predict dynamic attractors in combinatorial
  threshold-linear networks.
\newblock {\em Plos One}, 17(3):1--21, 03 2022.

\bibitem{lienkaemper2025}
C.~Lienkaemper and G.~Koch Ocker.
\newblock Diverse mean-field dynamics of clustered, inhibition-stabilized
  {{Hawkes}} networks via combinatorial threshold-linear networks.
\newblock 2025.
\newblock arXiv preprint arXiv:2506.06234.

\end{thebibliography}

% see TLN-domination-notes-7dec2024.tex for additional "dominoes" stuff that has been cut here.

\end{document}